\documentclass[aps,pre,twocolumn,groupedaddress]{revtex4-1}
\usepackage{graphicx}
\usepackage{dcolumn}
\usepackage{bm}
\usepackage{amsmath}
\usepackage{amssymb}
\usepackage{color}

\begin{document}

\title{Spatiotemporal pattern formation of membranes induced by surface molecular binding/unbinding}

\author{Hiroshi Noguchi}
\email[]{noguchi@issp.u-tokyo.ac.jp}
\affiliation{Institute for Solid State Physics, University of Tokyo, Kashiwa, Chiba 277-8581, Japan}


\begin{abstract}
Nonequilibrium membrane pattern formation is studied using meshless membrane simulation.
We consider that molecules bind to either surface of a bilayer membrane
and move to the opposite leaflet by flip--flop.
When binding does not modify the membrane properties and the transfer rates among the three states
are cyclically symmetric,
the membrane exhibits spiral-wave and homogeneous-cycling modes at high and low binding rates, respectively,
as in an off-lattice cyclic Potts model.
When binding changes the membrane spontaneous curvature,
these spatiotemporal dynamics are coupled with microphase separation.
When two symmetric membrane surfaces are in thermal equilibrium,
the membrane domains form 4.8.8 tiling patterns 
in addition to stripe and spot patterns.
In nonequilibrium conditions, moving biphasic domains and time-irreversible fluctuating patterns appear. The domains move ballistically or diffusively depending on the conditions.
\end{abstract}
\maketitle

\section{Introduction}

Various spatiotemporal patterns, such as traveling waves, have been observed in living cells.\cite{meri21,wu21,beta23,nogu24c,tani13,verg24}
Recent experiments have revealed that the oscillation of membrane shapes can be accompanied by wave propagation.\cite{lits18,wu18}
Membrane shape transformations are important processes in various functions, such as molecular transport, locomotion, and mitosis.\cite{mcma05,zimm06,suet14,kaks18,beth18,svit18,lutk12} 
For example, spherical buds are formed by curvature-inducing proteins such as clathrin and coat protein complexes in vesicle transports.\cite{kaks18,beth18,joha15,bran13,hurl10,mcma11}
The cell division position during mitosis is determined by the standing wave of Min proteins.\cite{lits18,lutk12,taka22}
The cell polarity of eukaryotic embryos is determined by a PAR protein pattern.\cite{meri21,lutk12,hoeg13}
Other membrane shape transformations can also be regulated by chemical waves.

The thermal equilibrium states of lipid membranes have been extensively studied and are sufficiently understood.
The morphology of liposomes and phase separation are well-understood based on the free energy, including bending energy.\cite{lipo95,seif97,svet14,honn09,baum03,veat03,yana08}
Curvature-inducing proteins bend the bound membrane and sense the membrane curvature (i.e., concentrate on the membranes of their preferred curvatures).\cite{baum11,has21,prev15,tsai21} The curvature dependencies of these proteins have been reproduced using mean-field theories.\cite{nogu22a,nogu23b,nogu24} 
The assembly of these proteins produces not only spherical buds and cylindrical membrane tubes\cite{baum11,mim12a,fros08,adam15,rama18,nogu16,nogu22b} but also periodic patterns such as hexagonal\cite{gout21} and checkerboard\cite{nogu23} arrays of curved domains and beaded-necklace-like membrane tubes.\cite{nogu21b}
In contrast to membranes in equilibrium, membranes in nonequilibrium have been much less explored.
The coupling of membrane deformation and reaction-diffusion dynamics has been simulated for traveling waves and Turing patterns.\cite{wu18,pele11,tame20,tame21,tame22,nogu23a}
On a flat surface, nucleation and growth can induce spatiotemporal patterns, such as the homogeneous-cycling (HC) and spiral-wave (SW) modes.\cite{nogu24a,nogu24b,nogu24d} 
However, the effects of thermal fluctuations on the spatiotemporal patterns of deformable membranes have not been investigated.

This study aims to clarify the spatiotemporal patterns of membranes caused by 
active phase separation with nucleation and growth.
We consider the binding/unbinding of molecules to/from both membrane surfaces. 
In our previous study,\cite{nogu23} we found stripe, checkerboard, and kagome-lattice patterns in thermal equilibrium.
In this study, we examined dynamic patterns in nonequilibrium conditions.
Nonequilibrium conditions can be generated by the differences in binding chemical potentials with the two surfaces
and active flip or flop by proteins (flippase or floppase, respectively).
We observed various patterns, such as spiral waves, moving biphasic domains, and time-irreversible fluctuations.

The simulation model and method are described in Sec.~\ref{sec:method}.
The results are presented and discussed in Sec.~\ref{sec:results}.
The binding/unbinding of molecules with zero and finite spontaneous curvatures are described 
in Sec.~\ref{sec:zero} and  Sec.~\ref{sec:cvs}, respectively.
Finally, a summary is presented in Sec.~\ref{sec:sum}.

\begin{figure}[t]
\includegraphics{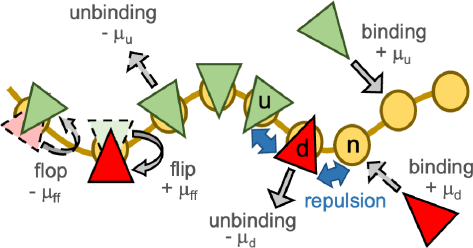}
\caption{
Schematic of the binding and unbinding of curvature-inducing molecules (proteins or surfactants)
that have a zero or finite spontaneous curvature.
The brown curve and yellow circles represent membrane and membrane binding sites, respectively.
The molecules (green and red triangles) bind to the membrane from the upper and lower solutions with
binding chemical potentials $\mu_{\mathrm{u}}$ and $\mu_{\mathrm{d}}$, respectively.
The molecules also move between the upper and lower membrane surfaces
with the flip--flop chemical potential $\mu_{\mathrm{ff}}$.
The three states of the membrane are represented by ``n'' (unbound), ``u'' (bound to the upper surface), and ``d'' (bound to the lower surface).
The solid and dashed gray arrows represent the dominant and secondary processes, respectively, in the nonequilibrium simulations.
}
\label{fig:cart}
\end{figure}

\section{Simulation Model and Method}\label{sec:method}

In a meshless membrane model,
membrane particles self-assemble into a one-layer sheet in a fluid phase.\cite{nogu09,nogu06,shib11,nogu19c}
Proteins or surfactant molecules bind to and unbind from the membrane.
Here, each membrane particle is a binding site and has three states: unbound membrane (bare, denoted by ``n''), bound to the upper surface of the membrane (denoted by ``u''), bound to the lower surface of the membrane (denoted by ``d''), as shown in Fig.~\ref{fig:cart}.
We call these membrane particles unbound and bound particles, respectively.
The particle diameter $\sigma$ is considered to be the size of a binding site, so that it is varied from $\sigma=5$ to $100$\,nm
depending on binding molecules.
The molecules bind to the upper and lower membrane surfaces with binding chemical potentials $\mu_{\mathrm{u}}$ and  $\mu_{\mathrm{d}}$, respectively.
The chemical potential for the molecular flip from the upper to the lower surface is  $\mu_{\mathrm{ff}}$, 
while that for flop from the lower to the upper surface is $-\mu_{\mathrm{ff}}$.
In thermal equilibrium, $\mu_{\mathrm{d}}=\mu_{\mathrm{u}}+\mu_{\mathrm{ff}}$.
The different binding chemical potentials ($\mu_{\mathrm{u}} \neq \mu_{\mathrm{d}}$)
can be generated by the concentration difference between the upper and the lower solutions.
Flip--flop is considered to occur spontaneously or actively via an enzyme through ATP hydrolysis.

The position and orientational vectors of the $i$-th particle are ${\bm{r}}_{i}$ and ${\bm{u}}_i$, respectively.
The membrane particles interact with each other via a potential $U=U_{\mathrm {rep}}+U_{\mathrm {att}}+U_{\mathrm {bend}}+U_{\mathrm {tilt}}+U_{\mathrm {pp}}$.
The potential $U_{\mathrm {rep}}$ is an excluded volume interaction with diameter $\sigma$ for all pairs of particles.
The solvent is implicitly accounted for by an effective attractive potential  as follows:
\begin{equation} \label{eq:U_att}
\frac{U_{\mathrm {att}}}{k_{\mathrm{B}}T} =  \frac{\varepsilon_{\mathrm{att}}}{4}\sum_{i} \ln[1+\exp\{-4(\rho_i-\rho^*)\}],
\end{equation}
where  $\rho_i= \sum_{j \ne i} f_{\mathrm {cut}}(r_{i,j})$,
 $\rho^*$ is the characteristic density with $\rho^*=7$, $\varepsilon_{\mathrm{att}}=8$
and $k_{\mathrm{B}}T$ is the thermal energy, as in our previous studies.\cite{gout21,nogu23}
$f_{\mathrm {cut}}(r)$ is a $C^{\infty}$ cutoff function\cite{nogu06}
 and $r_{i,j}=|{\bf r}_{i,j}|$, with ${\bf r}_{i,j}={\bf r}_{i}-{\bf r}_j$:
\begin{equation} \label{eq:cutoff}
f_{\mathrm {cut}}(r)=
\exp\Big\{b\Big(1+\frac{1}{(r/r_{\mathrm {cut}})^n -1}\Big)\Big\}\Theta(r_{\mathrm {cut}}-r),
\end{equation}
where $\Theta(x)$ is the unit step function, 
 $n=6$, $b=\ln(2) \{(r_{\mathrm {cut}}/r_{\mathrm {att}})^n-1\}$,
$r_{\mathrm {att}}= 1.9\sigma$, and $r_{\mathrm {cut}}=2.4\sigma$.
The bending and tilt potentials
are given by 
\begin{eqnarray} \label{eq:ubend}
\frac{U_{\mathrm {bend}}}{k_{\mathrm{B}}T} &=& \frac{k_{\mathrm {bend}}}{2} \sum_{i<j} ({\bm{u}}_{i} - {\bm{u}}_{j} - C_{\mathrm {bd}} \hat{\bm{r}}_{i,j} )^2 w_{\mathrm {cv}}(r_{i,j}), \\
\frac{U_{\mathrm {tilt}}}{k_{\mathrm{B}}T} &=& \frac{k_{\mathrm{tilt}}}{2} \sum_{i<j} [ ( {\bm{u}}_{i}\cdot \hat{\bm{r}}_{i,j})^2
 + ({\bm{u}}_{j}\cdot \hat{\bm{r}}_{i,j})^2  ] w_{\mathrm {cv}}(r_{i,j}),\  
\end{eqnarray}
where 
 $\hat{\bm{r}}_{i,j}={\bm{r}}_{i,j}/r_{i,j}$ and $w_{\mathrm {cv}}(r_{i,j})$ is a weight function. 
The spontaneous curvature is given by $C_0 = C_{\mathrm {bd}}/2\sigma$.\cite{shib11} 

The repulsive interactions of different states (u, d, and n) are represented by
\begin{equation} \label{eq:upp}
\frac{U_{\mathrm {pp}}}{k_{\mathrm{B}}T} =  \sum_{i,j} \varepsilon_{\mathrm {p},s_is_j} \exp\big( \frac{1}{r_{ij}/r_{\mathrm {cut}} -1} + b_0 \big) ,
\end{equation}
where $b_0= r_{\mathrm {cut}}/(r_{\mathrm {cut}}-\sigma)$.\cite{nogu12a}
At $r_{ij}=\sigma$, the potential height is $\varepsilon_{\mathrm {p},s_is_j}k_{\mathrm{B}}T$,
 where $s_i, s_j \in$ \{u, d, n\}.
In our previous study,\cite{nogu23} we used this repulsive interaction only between the bound particles (u and d).
In this study, we applied it for all three pairs (u--d, u--n, and d--n) to induce phase separation between all types of domain pairs.

The position ${\bf r}_{i}$ and 
 orientation ${\bf u}_{i}$ of membrane particles are updated by underdamped Langevin equations,
which are integrated by the leapfrog algorithm\cite{alle87,nogu11}
with $\Delta t=0.002\tau$. The time unit is $\tau= \sigma^2/D_0$,
where $D_0$ is the diffusion coefficient of the free membrane particles.
Two types of states [binding/unbinding (u vs. n and d vs. n) and flip--flop (u vs. d)] 
are stochastically switched by a Metropolis Monte Carlo (MC) procedure with the acceptance rate $p_{\mathrm {acpt}}$:
\begin{equation}\label{eq:Metro}
p_{\mathrm {acpt}} = \mathrm{min} \bigg[1, \exp\Big(\pm \frac{\Delta H}{k_{\mathrm{B}}T}\Big) \bigg]
\end{equation}
where the $+$ and $-$ signs refer to the forward and backward processes, respectively.
Here, $\Delta H= \Delta U - \mu_{\alpha}$ where $\Delta U$ is the energy difference between the two states
and $\mu_{\alpha}$ is the chemical potential of each process ($\alpha \in$ \{u, d, ff\}).
Thermal equilibrium states do not depend on the frequencies of these MC processes.
Conversely, in nonequilibrium, the dynamics can be varied by these frequencies.
For binding/unbinding and  flip--flop,
$\Gamma_{\mathrm {bind}}$ and  $\Gamma_{\mathrm {flip}}$ MC trials are performed, respectively, per particle at every MC step of $\tau_{\mathrm {MC}}=0.01\tau$.

\begin{figure}[t]
\includegraphics[]{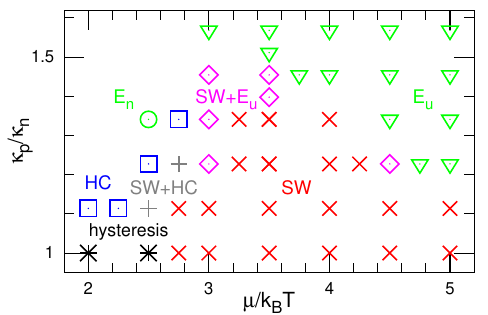}
\caption{
Phase diagram of the binding molecules of zero spontaneous curvatures ($C_0= 0$).
The (green) down-pointing triangles and circles represent the homogeneous phases of the upper-bound (E$_{\mathrm{u}}$) and unbound (E$_{\mathrm{n}}$) states, respectively.
The (red) crosses and (blue) squares represent the spiral-wave (SW) and homogeneous-cycling (HC) modes, respectively.
The (magenta) diamonds or (gray) pluses represent the temporal coexistence of SW and  E$_{\mathrm{u}}$ or HC, respectively.
The (black) asterisks represent the hysteresis of SW and HC, in which either mode appears depending on the initial state.
}
\label{fig:pdc0}
\end{figure}

\begin{figure}[t]
\includegraphics[]{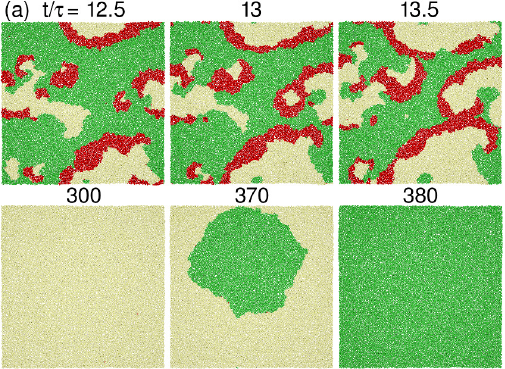}
\includegraphics[]{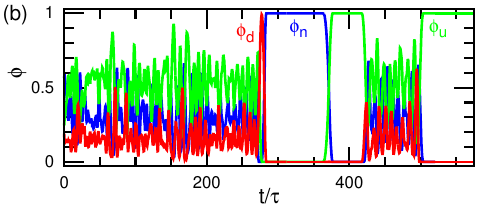}
\caption{
Temporal coexistence of the spiral-wave and homogeneous-cycling modes (SW+HC) for the binding molecules of zero spontaneous curvatures ($C_0= 0$)  at $\mu/k_{\mathrm {B}}T=2.75$ and $\kappa_{\mathrm{p}}/\kappa_{\mathrm{n}}=1.23$.
(a) Sequential snapshots. 
Green (medium gray) and red (dark gray) spheres represent the upper and lower bound particles (u and d), respectively.
Yellow (light gray) spheres represent unbound membrane particles (n).
The upper and lower snapshots show the spiral-wave and homogeneous-cycling modes, respectively.
(b) Time evolution of the densities of the three states.
}
\label{fig:tc0k12}
\end{figure}

\begin{figure}[t]
\includegraphics[]{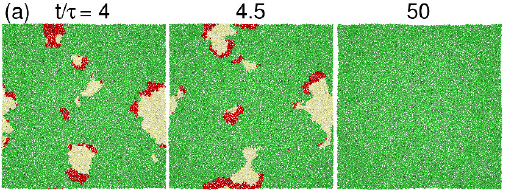}
\includegraphics[]{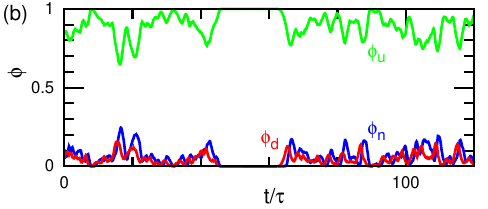}
\caption{
Temporal coexistence of spiral waves and homogeneous phases (SW+E$_{\mathrm{u}}$) for the binding molecules of zero spontaneous curvatures ($C_0= 0$)  at $\mu/k_{\mathrm {B}}T=3.5$ and $\kappa_{\mathrm{p}}/\kappa_{\mathrm{n}}=1.45$.
(a) Sequential snapshots. 
The first two snapshots show the spiral waves.
(b) Time evolution of the densities of the three states.
}
\label{fig:tc0k14}
\end{figure}

\begin{figure}[t]
\includegraphics[]{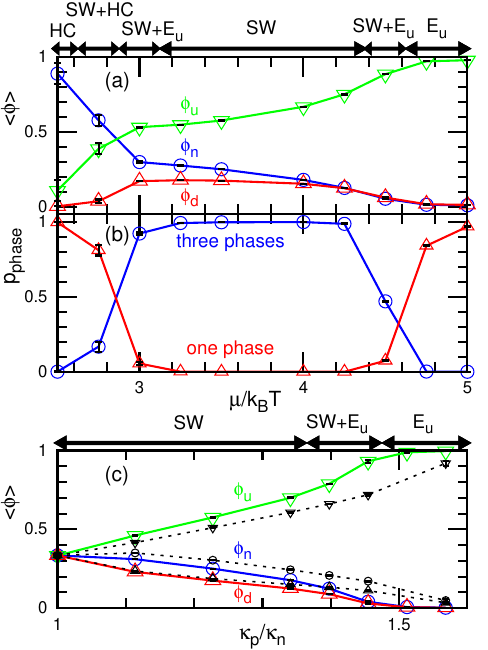}
\caption{
Dependence on $\mu$ and $\kappa_{\mathrm{p}}/\kappa_{\mathrm{n}}$
for the binding molecules of zero spontaneous curvatures ($C_0= 0$).
(a), (b) $\kappa_{\mathrm{p}}/\kappa_{\mathrm{n}}=1.23$. 
(c) $\mu/k_{\mathrm {B}}T=3.5$.
(a), (c) Mean densities $\langle \phi \rangle$ of the three states.
(b) Probabilities $p_{\mathrm{phase}}$ of a three-phase coexistence 
($\phi_{\mathrm{u}}>0.05$, $\phi_{\mathrm{d}}>0.05$, and $\phi_{\mathrm{n}}>0.05$) and one phase ($\phi_s>0.95$ with $s\in$ \{u, d, n\}).
The solid and dashed lines represent the data at $\Gamma_{\mathrm{MC}}=0.5$ and $0.005$, respectively.
The dynamic modes at $\Gamma_{\mathrm{MC}}=0.5$ are shown by the bidirectional arrows at the tops of (a) and (c).
}
\label{fig:c0km}
\end{figure}

A membrane consisting of $25~600$ membrane particles is simulated under periodic boundary conditions.
The $N\gamma T$ ensemble is used, where $N$ is the total number of particles and $\gamma$ is the surface tension,
so that the projected area of the square membrane thermally fluctuates and can be varied by the membrane deformation.\cite{fell95,nogu12}
Unless otherwise specified, we use a tension of $\gamma=0.5k_{\mathrm{B}}T/\sigma^2$, which is sufficiently large to prevent the budding of a bound domain.\cite{gout21,nogu23}
This tension corresponds to an average tension $\approx  0.02$\,mN/m, which is well below the usual lysis tension ($1$--$25$\,mN/m),\cite{evan00,evan03,ly04}
($\sigma\approx 10$\,nm and $k_{\mathrm{B}}T\approx 4\times10^{-21}$\,J).
For the unbound particles,
$C_0=0$ and $k_{\mathrm {bend}}=k_{\mathrm{tilt}}=10$ (corresponding bending rigidity $\kappa_{\mathrm {n}}/k_{\mathrm{B}}T=16.1 \pm 1$) are used.
The saddle-splay modulus $\bar{\kappa}$ is proportional to $\kappa$, as $\bar{\kappa}/\kappa=-0.9\pm 0.1$.\cite{nogu19}
These are typical values for lipid membranes.\cite{kara23,dimo14,hu12}

For the bound states, we examined two types of conditions:
1) no or slight changes of the membrane properties with a zero spontaneous curvature and
 2) generation of a finite spontaneous curvature.
In the first case, 
cyclically symmetric and quasi-symmetric conditions are considered.
The bound particles have no spontaneous curvature ($C_0=0$)
and identical or slightly high bending rigidities, $\kappa_{\mathrm {p}}$.
We use $\kappa_{\mathrm {p}}/\kappa_{\mathrm {n}}=1$--$1.6$ ($k_{\mathrm {bend}}=k_{\mathrm{tilt}}=10$--$15$),
since the bending rigidity linearly depends on $k_{\mathrm {bend}}=k_{\mathrm{tilt}}$ as $\kappa_{\mathrm {p}}/k_{\mathrm{B}}T=1.83k_{\mathrm {bend}}-2.2$.\cite{shib11}
The other parameters are symmetrically set as $\mu_{\mathrm{u}}=-\mu_{\mathrm{d}}=\mu_{\mathrm{ff}}=\mu$ and 
$\varepsilon_{\mathrm {p},\alpha\beta}=3$ for $\alpha,\beta\in$ \{u, d, n\},
and the MC rates are $\Gamma_{\mathrm {bind}}=\Gamma_{\mathrm {flip}}=\Gamma_{\mathrm {MC}}$.
We use $\Gamma_{\mathrm {MC}}=0.5$ unless  otherwise specified.
This model is an off-lattice version of the three-state cyclic Potts model.\cite{nogu24a,nogu24b}
Here, the particles (flipping sites) laterally diffuse, and their heights vertically fluctuate, unlike those in the lattice Potts models. Note that this cyclic condition can be alternatively realized by the binding only on one leaflet of the bilayer and chemical reaction instead of the flip--flop.\cite{nogu24a}

In the second case, the binding of curvature-inducing molecules (proteins or surfactants) is considered.
The molecule bound to the upper and lower surfaces induces positive and negative spontaneous curvatures of $C_0\sigma=\pm 0.1$, respectively,
and their bending rigidities are high as $\kappa_{\mathrm {u}}/k_{\mathrm{B}}T=\kappa_{\mathrm {d}}/k_{\mathrm{B}}T=144 \pm 7$ ($k_{\mathrm {bend}}=k_{\mathrm{tilt}}=80$).
We use $\varepsilon_{\mathrm {p,ud}}=3$, $\varepsilon_{\mathrm {p,un}}=\varepsilon_{\mathrm {p,dn}}=2$,
 $\Gamma_{\mathrm {bind}}= 1$, and $\Gamma_{\mathrm {flip}}=0.01$.
This parameter set is identical to that in our previous study,\cite{nogu23} except for the repulsion strength ($\varepsilon_{\mathrm {p,un}}=\varepsilon_{\mathrm {p,dn}}=0$ in Ref.~\citenum{nogu23}) for bound and unbound particles.
In Ref.~\citenum{nogu23}, we obtained only steady spatial patterns in nonequilibrium.
A sufficiently large repulsion to form three types of domains is key to obtaining spatiotemporal patterns.
The equilibrium membrane behaviors of membranes without molecular binding, with binding to the upper surface, and binding to both surfaces are described in Refs.~\citenum{shib11}, \citenum{gout21}, and \citenum{nogu23}, respectively.

The mean cluster sizes are calculated to characterize various phases. 
Two membrane particles of each state are considered to belong to the same cluster
when the inter-particle distance
is less than $r_{\mathrm {att}}$. 
The mean size of the clusters is given by
$N_{s,\mathrm {cl}}= (\sum_{i_{s,{\mathrm {cl}}}=1}^{N_s} i_{s,\mathrm{cl}}^2 n^{s,{\mathrm {cl}}}_i)/N_s$,
where $n^{s,{\mathrm {cl}}}_i$  is the number of clusters of size $i_{s,\mathrm{cl}}$ 
and $N_s$ is the total number of each state ($s \in$ \{u, d, n\}).
The mean cluster size of  each state
is normalized by the mean total number as $\chi_{s}=\langle N_{s,\mathrm{cl}}\rangle/\langle N_{s}\rangle$. 
A large percolated cluster results in $\chi \simeq 1$.
The vertical span of the membrane is calculated from 
the membrane height variance as 
$z_{\mathrm {mb}}^2=\sum_{i}^{N} (z_i-z_{\mathrm G})^2/N$,
where $z_{\mathrm G}=\sum_{i}^{N} z_i/N$. 
The statistical errors are calculated from three or more independent runs.

\section{Simulation Results}\label{sec:results}

\subsection{Membrane with Zero-Spontaneous-Curvature}\label{sec:zero}

First, we describe the pattern formation of membranes with a zero spontaneous curvature.
The cyclic dynamics reported in the square-lattice Potts model\cite{nogu24a,nogu24b}
are also obtained in this off-lattice model (see Figs.~\ref{fig:pdc0}--\ref{fig:c0km}).
Three dynamic modes (SW, HC, and steady one-phase mode (E$_s$)) and their temporal coexistence are obtained (see the phase diagram shown in Fig.~\ref{fig:pdc0}).
The temporal coexistence of SW and HC (SW+HC) is shown in Fig.~\ref{fig:tc0k12},
where ``+'' represents the temporal coexistence.
In SW mode, the boundary of the u and d phases (or the d and n and the n and u phases) moves 
in the direction of the u phase (or d and n phases, respectively), and
the contact point of the three phases becomes the center of a spiral wave 
(see the upper three snapshots in Fig.~\ref{fig:tc0k12}(a) and Movie~S1).
In HC mode, the membrane is dominantly occupied by one of the three phases for most of the period, 
and the dominant phases change cyclically as $s= \mathrm{n}\to \mathrm{u} \to \mathrm{d} \to \mathrm{n}$.
These phase changes are caused by nucleation and growth, as shown in the lower three snapshots in Fig.~\ref{fig:tc0k12}(a).

The mean densities $\langle\phi\rangle$ 
of the three states are identical ($=1/3$) in the symmetric condition ($\kappa_{\mathrm{p}}/\kappa_{\mathrm{n}}=1$).
At low $\mu$, either the HC or SW mode appears depending on the initial state owing to hysteresis;
 the SW mode only appears at high $\mu$ (see the bottom of Fig.~\ref{fig:pdc0}).
This phenomenon corresponds to the discontinuous transition obtained in large systems ($N> 37000$) of the lattice Potts model.\cite{nogu24a} We also confirmed that this transition occurs in tensionless membranes ($\gamma=0$).

As $\kappa_{\mathrm{p}}$ increases, $\langle\phi_{\mathrm{u}}\rangle$ increases.
Subsequently, the u domain covers the whole membrane (E$_{\mathrm{u}}$ mode) 
(see Figs.~\ref{fig:tc0k14} and \ref{fig:c0km}(c), and Movie~S2).
This E$_s$ mode is similar to a homogeneous phase in thermal equilibrium.
This change is caused by the mechanism reported for the lattice Potts model.\cite{nogu24b}
At high $\kappa_{\mathrm{p}}$, the unbinding is promoted since the unbound membrane has a larger undulation than the bound membrane (i.e, the unbinding gains an entropy).
Hence, the duration of the d-dominant phase is reduced. This enhancement of the unbinding also suppresses the nucleation of d domains in the u-dominant phase,
since small d domains disappear through the two-step process of $\mathrm{d}\to \mathrm{n} \to \mathrm{u}$.\cite{nogu24b}
Consequently, the duration of the E$_\mathrm{u}$ mode increases.
When the MC rates of binding/unbinding/flip--flop are reduced to $\Gamma_{\mathrm{MC}}=0.005$,
the effects of  $\kappa_{\mathrm{p}}$ are weakened, since the membrane can relax to the preferred projected area for the local bending rigidity (compare the dashed and solid lines in Fig.~\ref{fig:c0km}(c)).

The modes are determined according to the manner used in Refs.~\citenum{nogu24a} and \citenum{nogu24b} as follows.
When the probability of three-phase coexistence ($\langle\phi_{\mathrm{u}}\rangle>0.05$, $\langle\phi_{\mathrm{d}}\rangle>0.05$, and $\langle\phi_{\mathrm{n}}\rangle>0.05$) is greater than $0.05$, the SW mode exists (see the blue lines with circles in Fig.~\ref{fig:c0km}(b)).
When that of the one-phase state ($\langle\phi_s\rangle>0.95$ with $s\in$ \{u, d, n\}) is greater than $0.05$,
the HC or E$_s$ mode exists. In the HC mode, all three states have a peak at $\phi_s\approx 1$.
In the E$_s$ mode, the $s$ state has the highest peak at $\phi_s\approx 1$, and at least one of the other states does not have this peak.
In the upper-right region of the phase diagram (Fig.~\ref{fig:pdc0}), the E$_{\mathrm{u}}$ mode is formed.
At $\kappa_{\mathrm{p}}/\kappa_{\mathrm{n}}=1.23$,
the mode changes from HC to SW and E$_{\mathrm{u}}$ with increasing $\mu$ through the coexistence of modes (SW+HC and SW+E$_{\mathrm{u}}$), as shown in Fig.~\ref{fig:c0km}.
Hence, the discontinuous transition between the SW and HC modes becomes continuous at $\kappa_{\mathrm{p}}/\kappa_{\mathrm{n}}\simeq 1.2$.

Here, the SW mode is obtained only at  $\kappa_{\mathrm{p}}/\kappa_{\mathrm{n}}<1.5$ while keeping the other properties.
Therefore, a spiral wave spreading over a large membrane area is difficult to obtain when the binding largely modifies 
the membrane properties and/or strong interactions between the bound molecules.
However, biphasic domains (Fig.~\ref{fig:tc0k14}(a)) can migrate under the conditions described in Sec.~\ref{sec:cvneq}.

\subsection{Binding/Unbinding of Curvature-Inducing Molecules}\label{sec:cvs}

\subsubsection{Equilibrium Phases}\label{sec:cveq}

\begin{figure}[t]
\includegraphics[]{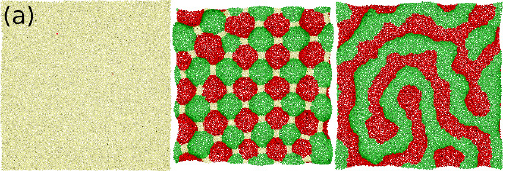}
\includegraphics[]{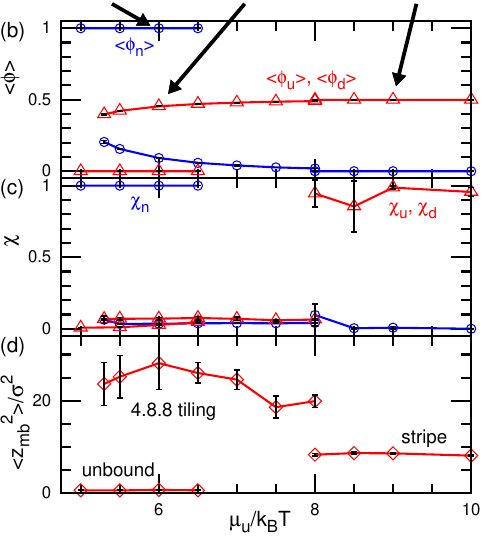}
\caption{
Equilibrium states for the binding molecules of finite curvatures ($C_0\sigma= \pm 0.1$) in up--down symmetrical conditions ($\mu_{\mathrm{u}}=\mu_{\mathrm{d}}$ and $\mu_{\mathrm{ff}}=0$) 
(a) Snapshots. From left to right, an unbound membrane at  $\mu_{\mathrm{u}}/k_{\mathrm {B}}T=6$, 
 4.8.8-tiling domains at  $\mu_{\mathrm{u}}/k_{\mathrm {B}}T=6$, and stripe domains at $\mu_{\mathrm{u}}/k_{\mathrm {B}}T=9$.
(b) Mean densities of bound molecules $\langle\phi_{\mathrm{u}}\rangle=\langle\phi_{\mathrm{d}}\rangle$ and the unbound membrane $\langle\phi_{\mathrm{n}}\rangle$.
(c) Mean cluster sizes normalized by the total number of each type of particle:
for the u state, $\chi_{\mathrm{u}}=\langle N_{\mathrm{ u,cl}}\rangle/\langle N_{\mathrm{u}}\rangle$;
for the n state (unbound), $\chi_{\mathrm {n}}=\langle N_{\mathrm {n,cl}}\rangle/\langle N_{\mathrm{n}}\rangle$. 
(d) Membrane vertical span $z_{\mathrm {mb}}$.
The transition between the unbound and 4.8.8-tiling phases
is first order, and two phases coexist as (meta-)stable states around the transition point.
}
\label{fig:eq}
\end{figure}

Before discussing nonequilibrium dynamics,
 we briefly describe the phase behaviors under up--down symmetrical conditions in thermal equilibrium ($\mu_{\mathrm{u}}=\mu_{\mathrm{d}}$ and $\mu_{\mathrm{ff}}=0$)
for curvature-inducing molecules of $C_0\sigma=\pm 0.1$.
The molecules bind to the upper and lower surfaces with the same amount, as shown in Fig.~\ref{fig:eq}(a) and (b).
As the binding chemical potentials ($\mu_{\mathrm{u}}=\mu_{\mathrm{d}}$) increase,
the densities of the molecules ($\phi_{\mathrm{u}}=\phi_{\mathrm{d}}$) increase,
and eventually, the membrane is almost completely covered by the molecules ($\langle\phi_{\mathrm{u}}\rangle=\langle\phi_{\mathrm{d}}\rangle \approx  0.5$).
Thus, the membrane forms a stripe of the u and d domains at $\mu_{\mathrm{u}}/k_{\mathrm{B}}T \geq 8$ (see the right snapshot in Fig.~\ref{fig:eq}(a)), in which
most bound u (d) particles belong to the largest u (d) domain, 
that is, $\chi_{\mathrm{u}}=\chi_{\mathrm{d}}\simeq 1$ (see Fig.~\ref{fig:eq}(c)).

\begin{figure}[]
\includegraphics[]{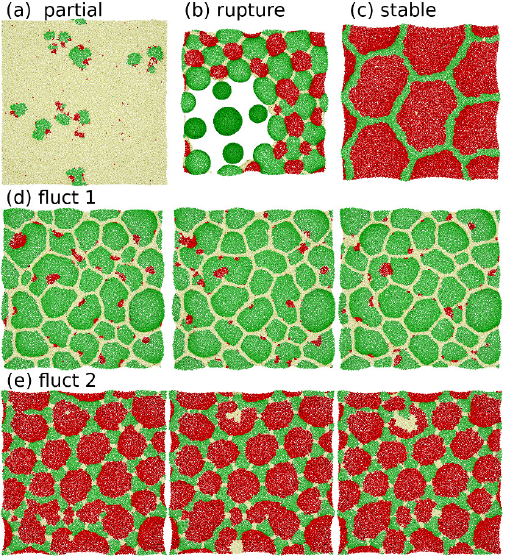}
\includegraphics[]{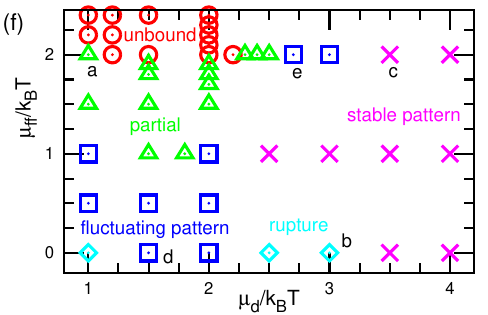}
\caption{
Nonequilibrium phases for the binding molecules of finite curvatures ($C_0\sigma= \pm 0.1$)  at  $\mu_{\mathrm{u}}/k_{\mathrm {B}}T=10$. 
(a)--(e) Snapshots. 
(a) Partial pattern at  $\mu_{\mathrm{d}}/k_{\mathrm {B}}T=1$ and $\mu_{\mathrm{ff}}/k_{\mathrm {B}}T=2$. 
(b) Rupture at  $\mu_{\mathrm{d}}/k_{\mathrm {B}}T=3$ and $\mu_{\mathrm{ff}}=0$. 
(c) Stable pattern at  $\mu_{\mathrm{d}}/k_{\mathrm {B}}T=3.5$ and $\mu_{\mathrm{ff}}/k_{\mathrm {B}}T=2$. 
(d) Fluctuating pattern at  $\mu_{\mathrm{d}}/k_{\mathrm {B}}T=1.5$ and $\mu_{\mathrm{ff}}=0$. The time interval of the sequential snapshots is $100\tau$.
(e) Fluctuating pattern at  $\mu_{\mathrm{d}}/k_{\mathrm {B}}T=2.7$ and $\mu_{\mathrm{ff}}/k_{\mathrm {B}}T=2$. 
The time interval is $60\tau$.
(f) Phase diagram.
The circles, triangles, squares, crosses, and diamonds represent
the unbound membranes, partial patterns, fluctuating patterns, stable patterns, and ruptures, respectively.
The letters (a--e) indicate the corresponding data points of the snapshots in (a--e).
}
\label{fig:cp10}
\end{figure}

At medium strengths of the binding potential ($5.3 \leq \mu_{\mathrm{u}}/k_{\mathrm{B}}T \leq 8$),
the u and d domains form a square lattice,
and the vertices are stabilized by the n domains (see the middle snapshot in Fig.~\ref{fig:eq}(a)).
Owing to the symmetry,  an even number of these bound domains surround the vertex.
Since we added repulsion between the bound and unbound domains,
the n domains at the vertices are larger than those in our previous study.\cite{nogu23}
Thus, the bound (u and d) domains are shaped as octagons and the n domain is shaped as a square.
When the bound domains form a regular octagon,
this domain pattern corresponds to the semi-regular tessellation 4.8.8, which consists of regular octagons and squares;
this pattern is one of 
eight tilings to cover a flat plane by two or more regular polygons.
Hence, we call this state 4.8.8 tiling.
Since the u and d domains
bend in the positive and negative directions, respectively,
the bound membrane exhibits a bumped stripe or spherical-cap shape (see Fig.~\ref{fig:eq}(d)).

The transitions between the unbound and 4.8.8 tiling phases
are discontinuous, and the metastable states exist around the transition point (see Fig.~\ref{fig:eq}(b)--(d)).
When the binding potential exceeds $\mu_{\mathrm{u}}/k_{\mathrm{B}}T = 8$, 
the tiling membrane is ruptured from the boundaries of the n domains.
This is due to the addition of the repulsion between the bound and unbound particles in the present study,
since the discontinuous transition is obtained without rupture in the absence of this repulsion.\cite{nogu23}
In asymmetric conditions ($\mu_{\mathrm{u}}\ne \mu_{\mathrm{d}}$ and $\mu_{\mathrm{d}}=\mu_{\mathrm{u}}+\mu_{\mathrm{ff}}$), kagome-lattice domain patterns reported in Ref~\citenum{nogu23} is also obtained in the present system.

\begin{figure}[t]
\includegraphics[]{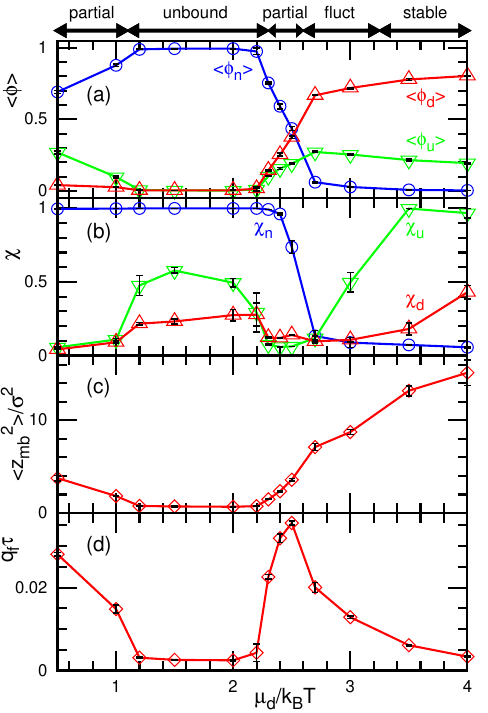}
\caption{
Dependence on the binding chemical potential $\mu_{\mathrm{d}}$
 for the binding molecules of finite curvatures ($C_0\sigma= \pm 0.1$) 
at  $\mu_{\mathrm{u}}/k_{\mathrm {B}}T=10$ and  $\mu_{\mathrm{ff}}/k_{\mathrm {B}}T=2$.
(a) Mean densities of the three states.
(b) Normalized mean cluster sizes $\chi_{\mathrm{u}}$, $\chi_{\mathrm{d}}$,  and $\chi_{\mathrm{n}}$.
(c) Membrane vertical span $z_{\mathrm {mb}}$.
(d) Flow rate $q_{\mathrm {f}}$ of the binding molecules from the upper to the lower solution.
The dynamic modes are indicated by the bidirectional arrows at the top of the figure.
}
\label{fig:cp10e2}
\end{figure}

\begin{figure}[t]
\includegraphics[]{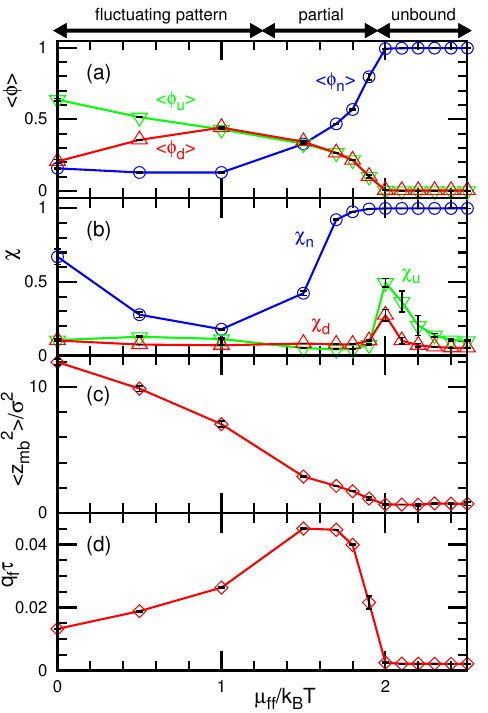}
\caption{
Dependence on the flip chemical potential $\mu_{\mathrm{ff}}$
 for the binding molecules of finite curvatures ($C_0\sigma= \pm 0.1$) 
at  $\mu_{\mathrm{u}}/k_{\mathrm {B}}T=10$ and  $\mu_{\mathrm{d}}/k_{\mathrm {B}}T=2$.
(a) Mean densities of the three states.
(b) Normalized mean cluster sizes $\chi_{\mathrm{u}}$, $\chi_{\mathrm{d}}$,  and $\chi_{\mathrm{n}}$.
(c) Membrane vertical span $z_{\mathrm {mb}}$.
(d) Flow rate $q_{\mathrm {f}}$ of the binding molecules from the upper to the lower solution.
The dynamic modes are indicated by the bidirectional arrows at the top of the figure.
}
\label{fig:cp10d2}
\end{figure}

\begin{figure}[t]
\includegraphics[]{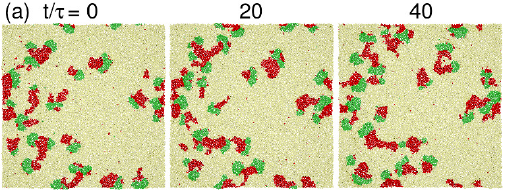}
\includegraphics[]{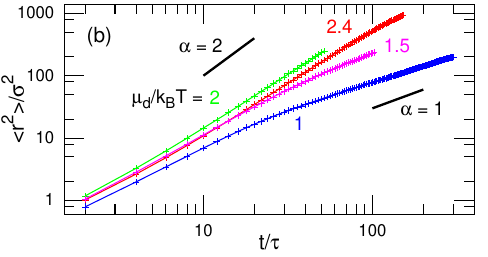}
\caption{
Motion of the biphasic domains of finite curvatures ($C_0\sigma= \pm 0.1$)  at  $\mu_{\mathrm{u}}/k_{\mathrm {B}}T=10$ and  $\mu_{\mathrm{ff}}/k_{\mathrm {B}}T=2$.
(a) Sequential snapshots with an interval of $20\tau$ at  $\mu_{\mathrm{d}}/k_{\mathrm {B}}T=2.3$. 
(b) Mean squared distance $\langle r(t)^2 \rangle$ of the center of domain u at $\mu_{\mathrm{d}}/k_{\mathrm {B}}T=1$, $1.5$, $2$, and $2.4$.
The two black straight lines represent exponents $\alpha=1$ and $2$.
}
\label{fig:rsd}
\end{figure}

\begin{figure}[t]
\includegraphics[]{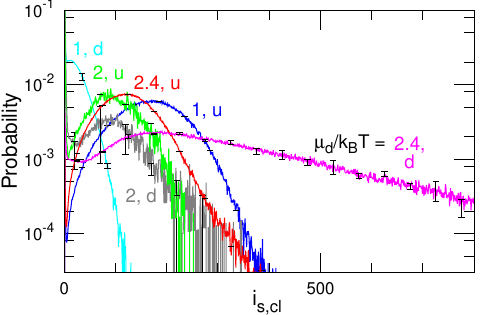}
\caption{
Size distribution of the u and d domains for $\mu_{\mathrm{d}}/k_{\mathrm {B}}T=1$, $2$, and $2.4$ at $\mu_{\mathrm{u}}/k_{\mathrm {B}}T=10$ and  $\mu_{\mathrm{ff}}/k_{\mathrm {B}}T=2$ (corresponding to the data shown in Fig.~\ref{fig:rsd}).
The d domains become larger with increasing $\mu_{\mathrm{d}}$.
}
\label{fig:cnhis}
\end{figure}

\subsubsection{Nonequilibrium Patterns}\label{sec:cvneq}

Next, we consider a membrane far from equilibrium.
Although this can be done in several ways,
we select a high binding potential on the upper surface ($\mu_{\mathrm{u}}/k_{\mathrm{B}}T = 10$),
 a low binding potential on the lower surface ($0\leq \mu_{\mathrm{d}}/k_{\mathrm{B}}T \leq 4$),
and a low flip potential ($0\leq \mu_{\mathrm{d}}/k_{\mathrm{B}}T \leq 2.5$).
Hence, the molecules bind to the upper surface but transfer to the lower surface and unbind in the lower solution (see Fig.~\ref{fig:cart}).
This cyclic sequence is identical to that in the flat membrane described in Sec.~\ref{sec:zero}.
However, they are coupled with the microphase separation described in Sec.~\ref{sec:cveq}.

The dynamic phase diagram is shown in Fig.~\ref{fig:cp10}.
At a relatively large $\mu_{\mathrm{d}}$ (i.e., a small difference between $\mu_{\mathrm{d}}$ and $\mu_{\mathrm{u}}$),
 stable patterns are formed, similar to those reported in our previous study.\cite{nogu23}
A hexagonal pattern consisting of large hexagonal d domains and the percolated hexagonal network of a u domain
is formed at $\mu_{\mathrm{d}}/k_{\mathrm{B}}T \simeq 4$ and $\mu_{\mathrm{ff}}/k_{\mathrm{B}}T \simeq 2$,
as shown in Fig.~\ref{fig:cp10}(c).
Hence, the d domains have a large area fraction $\langle\phi_{\mathrm{d}}\rangle \simeq 0.8$, 
the normalized mean cluster size of the domain is $\chi_{\mathrm{u}} \approx 1$,
and the bumped shape causes a large vertical span $z_{\mathrm {mb}}$ (see Fig.~\ref{fig:cp10e2}(a)--(c)).
At a low flip potential $\mu_{\mathrm{ff}}\simeq 0$, 
the 4.8.8 tiling occurs. As $\mu_{\mathrm{d}}$ decreases, the membrane is ruptured, leading to vesicle formation,
as shown in Fig.~\ref{fig:cp10}(b).

At a slightly lower $\mu_{\mathrm{d}}$ than that in the stable-pattern region, 
it is found that the domain patterns exhibit time-irreversible fluctuations (called fluctuating patterns).
At $\mu_{\mathrm{d}}/k_{\mathrm{B}}T = 1.5$ and $\mu_{\mathrm{ff}}=0$,
the u and n domains form a hexagonal pattern, but
a small d domain moves through a u domain, accompanied by a narrow n domain behind it (see Fig.~\ref{fig:cp10}(d) and Movie~S3).
The backward process is negligible owing to the cyclic relation of the chemical potentials.
This time-irreversible motion of d domains generates frequent changes in local patterns.
At $\mu_{\mathrm{d}}/k_{\mathrm{B}}T \simeq 3$ and $\mu_{\mathrm{ff}}/k_{\mathrm{B}}T=2$,
an n domain appears inside of a d domain and spreads, but the resultant large n domain is rapidly replaced by u and d domains (see Fig.~\ref{fig:cp10}(e) and Movie~S4).
In both cases, the domain-spreading process obtained in the flat membrane (Sec.~\ref{sec:zero})
occurs only locally in the separated space owing to the microphase separation.
Hence, the wave propagation is restricted to the inside of the domains but frequently modifies the local domain patterns. 
In the fluctuating patterns, the percolation of network domains is often broken
so that the membranes have low values of $\chi_{\mathrm{u}}$ (see Figs.~\ref{fig:cp10e2}(b) and \ref{fig:cp10d2}(b)).
In the pattern fluctuations, the domain size largely fluctuates (see Fig.~\ref{fig:cp10}(d)),
and at $\mu_{\mathrm{d}}/k_{\mathrm{B}}T=1$ and $\mu_{\mathrm{ff}}=0$ the formation of large spherical buds results in membrane rupture 
(the diamond at the bottom left in Fig.~\ref{fig:cp10}(f)).

As $\mu_{\mathrm{d}}$ further decreases at $\mu_{\mathrm{ff}}/k_{\mathrm{B}}T=2$,
most of the membrane area is unbound, and isolated biphasic (u and d) domains appear (see Figs.~\ref{fig:cp10}(a) and \ref{fig:rsd}(a) and Movies~S5 and S6). 
We distinguish whether the membrane is unbound or partially forms domains (called partial pattern) by the area fraction of the unbound state:
unbound membranes at $\langle\phi_{\mathrm{n}}\rangle > 0.95$ and partial patterns at  $\langle\phi_{\mathrm{n}}\rangle \leq 0.95$ (see Figs.~\ref{fig:cp10e2}(a) and \ref{fig:cp10d2}(a)).
Biphasic domains are occasionally formed in the unbound membranes,
whereas they appear most of the time in the partial patterns.
The molecular transfer from the upper to the lower solution occurs more frequently in the partial and fluctuating patterns
(see Figs.~\ref{fig:cp10e2}(d) and \ref{fig:cp10d2}(d)).

Last, we analyze the motion of the biphasic domains in the regions of the unbound membrane and partial pattern. 
The domain consisting of two (u and d) domains with equal size moves ballistically in the direction of the u domain, as shown in Fig.~\ref{fig:rsd}(a) and Movie~S5.
Conversely, when the d domains are much smaller than the u domain,
the domain moves diffusively (see Fig.~\ref{fig:cp10}(a) and Movie~S6).
This is due to frequent changes in the positions of the d domains.
These moving biphasic domains have a narrower size distribution than those in the flat membrane (see Fig.~\ref{fig:cnhis}) and
thus, have sufficiently long lifetimes for quantitative analysis.

Figure~\ref{fig:rsd}(b) shows
 the mean squared displacement $\langle r(t)^2\rangle$ of the domain center
 for a u domain of cluster size $i_{u,\mathrm{cl}}>50$.
The displacement of each cluster is calculated every $2\tau$
if both differences of the displacement and cluster size from those of the previous step 
are small ($\Delta r^2 < (5\sigma)^2$ and $\Delta i_{u,\mathrm{cl}} < 50$)
in order to exclude coalesced and divided clusters.
Ballistic and diffusive motions exhibit $\langle r^2\rangle\propto t^\alpha$ 
with exponents of $\alpha=2$ and $1$, respectively.
For $\mu_{\mathrm{d}}/k_{\mathrm{B}}T \geq 2$, $\alpha=2$
is obtained, i.e., it is ballistic.
The biphasic domain has a maximum velocity when it consists of u and d domains with equal size at $\mu_{\mathrm{d}}/k_{\mathrm{B}}T=2$  (see Figs.~\ref{fig:rsd}(b) and \ref{fig:cnhis}).
For $\mu_{\mathrm{d}}/k_{\mathrm{B}}T = 1.5$, the exponent changes from $\alpha=2$ to $1$ at $t/\tau \simeq 30$,
such that the ballistic domain motion becomes diffusive.
For $\mu_{\mathrm{d}}/k_{\mathrm{B}}T = 1$, $\alpha\sim 1$ even at $t/\tau < 30$.
The division and disappearance of d domains connecting the u domain change the moving direction 
so that the motion becomes diffusive.
A low exponent $\alpha<1$ at $t/\tau \simeq 100$ for $\mu_{\mathrm{d}}/k_{\mathrm{B}}T = 1$
may suggest that fast-moving domains coalesce or divide earlier than slow-moving domains.
The number of remaining domains decreases exponentially over time.

\section{Summary}\label{sec:sum}

We have studied the spatiotemporal patterns of membranes with three states.
When the three states are identical but separate from each other,
the membrane exhibits spiral waves of domains of three states at the cyclic flipping condition as an off-lattice cyclic Potts model.
At low flipping energies, the cycling of three homogeneous phases also occurs depending on the initial state.
When the bending rigidities of the three states are different, the temporal coexistence of these two modes and the non-cyclic one-phase mode also appear, since the membrane undulation modifies the stability of the states.

Next, we considered that molecules (e.g., surfactants and proteins) with positive spontaneous curvatures bind to and unbind from two surfaces of the membranes and move between the two surfaces by flip--flop.
In thermal equilibrium conditions, strip and 4.8.8 tiling patterns are formed.
In nonequilibrium conditions, the hexagonal patterns exhibit time-irreversible fluctuations in which small domains move ballistically in the large domains of other states, 
and the domain patterns largely fluctuate.
Moreover, biphasic domains move ballistically or diffusively depending on the conditions.
These moving biphasic domains have a narrow size distribution and longer lifetime than those in the flat membrane.

In the present system, a nonequilibrium condition can be generated by the difference in chemical potential between the upper and lower solutions.
Since the chemical potential can be varied by the concentration ($\Delta \mu = \Delta\rho k_{\mathrm{B}}T$ in dilute solutions), 
it is easily set up in experiments.
Experimentally, 
Miele and co-workers reported the shape transitions of the vesicles 
induced by the binding, unbinding, and flip--flop of surfactant molecules from the outer to the inner solutions.\cite{miel20,holl21}
Although the membrane is homogeneous in their experiments,
phase separation can be induced by using two types of surfactants, in which the hydrophobic segments consist of alkyl or fluoroalkyl chains 
because these chains can exhibit phase separation.\cite{xiao16,waka22}
Thus, the dynamic patterns obtained in the present study can be experimentally examined.

\begin{acknowledgments}
We thank Jean-Baptiste Fournier for stimulating discussion.
This work was supported by JSPS KAKENHI Grant Number JP24K06973. . 
\end{acknowledgments}


%

\end{document}